\documentclass[
 reprint,
superscriptaddress,
 amsmath,amssymb,
 aps,
superscriptaddress,
 amsmath,amssymb,
 aps,longbibliography
]{revtex4-1}

\usepackage{graphicx}
\usepackage{dcolumn}
\usepackage{bm}
\usepackage{amsmath,amssymb,amstext,amsthm}
\usepackage{braket}
\usepackage{bbold}
\usepackage{bbm}
\usepackage{xcolor}
\usepackage{ gensymb }

\begin{document}

\preprint{APS/123-QED}
\setlength{\abovedisplayskip}{1pt}
\title{Human psychophysical discrimination of spatially dependant Pancharatnam-Berry phases in optical spin-orbit states}

\author{D. Sarenac}
\affiliation{Institute for Quantum Computing, University of Waterloo,  Waterloo, ON, Canada, N2L3G1}
\author{A. E. Silva} 
\affiliation{School of Optometry and Vision Science, University of Waterloo, Waterloo, ON, Canada, N2L3G1}
\author{C. Kapahi} 
\affiliation{Institute for Quantum Computing, University of Waterloo,  Waterloo, ON, Canada, N2L3G1}
\affiliation{Department of Physics, University of Waterloo, Waterloo, ON, Canada, N2L3G1}
\author{D. G. Cory}
\affiliation{Institute for Quantum Computing, University of Waterloo,  Waterloo, ON, Canada, N2L3G1}
\affiliation{Department of Chemistry, University of Waterloo, Waterloo, ON, Canada, N2L3G1}
\author{B. Thompson} 
\affiliation{School of Optometry and Vision Science, University of Waterloo, Waterloo, ON, Canada, N2L3G1}
\affiliation{Centre for Eye and Vision Research, Hong Kong}
\author{D. A. Pushin}
\affiliation{Institute for Quantum Computing, University of Waterloo,  Waterloo, ON, Canada, N2L3G1}
\affiliation{Department of Physics, University of Waterloo, Waterloo, ON, Canada, N2L3G1}

\date{\today}


\pacs{Valid PACS appear here}


\begin{abstract}

We tested the ability of human observers to discriminate distinct profiles of spatially dependant geometric phases when directly viewing stationary structured light beams. Participants viewed polarization coupled orbital angular momentum (OAM) states, or ``spin-orbit'' states, in which the OAM was induced through Pancharatnam-Berry phases. The coupling between polarization and OAM in these beams manifests as spatially dependant polarization. Regions of uniform polarization are perceived as specifically oriented Haidinger's brushes, and study participants discriminated between two spin-orbit states based on the rotational symmetry in the spatial orientations of these brushes. Participants used self-generated eye movements to prevent adaptation to the visual stimuli. After initial training, the participants were able to correctly discriminate between two spin-orbit states, differentiated by OAM $=\pm1$, with an average success probability of $69\%$ ($S.D. = 22\%$, $p = 0.013$). These results support our previous observation that human observers can directly perceive spin-orbit states, and extend this finding to non-rotating beams, OAM modes induced via Pancharatnam-Berry phases, and the discrimination of states that are differentiated by OAM.

\end{abstract}
\maketitle

The phase that is acquired during cyclic evolution governed by a slow change of parameters is known as the geometric phase~\cite{cohen2019geometric}. The Aharonov-Bohm phase in quantum mechanics~\cite{aharonov1959significance} and the Pancharatnam-Berry phase in optics~\cite{pancharatnam1956generalized,berry1987adiabatic} are the two of the most well-known examples of phase shifts with geometrical origin, and they have had a profound impact on a wide range of areas in physics~\cite{wilczek1989geometric,berry1990anticipations,bachtold1999aharonov,noguchi2014aharonov}. Unlike typical phase shifts that arise from optical path differences, the Pancharatnam-Berry phase is induced when the polarization state traces out a geodesic triangle on the Poincar\'e sphere~\cite{pancharatnam1956generalized,berry1987adiabatic}. It has led to the development of optical components that enable polarization dependant wavefront shaping and novel methods of inducing polarization coupled orbital angular momentum (OAM) through spin-orbit coupling~\cite{biener2002formation,hasman2003polarization,bomzon2002space,marrucci2006optical,nsofini2016spin,sarenac2019generation}. Optical waves carrying OAM possess a helical wavefront and are described by the phase term $e^{i\ell \phi}$, where $\phi$ is the azimuthal coordinate and $\ell$ is the OAM value~\cite{LesAllen1992}. Beams with polarization coupled OAM, also known as ``vector vortex beams'' or ``spin-orbit'' beams, may be prepared to be non-separable in polarization and spatial modes~\cite{Konrad2019QuantumLight,mclaren2015measuring} and manifest dynamic 2D polarization topologies ~\cite{zhan2009cylindrical,cardano2012polarization,galvez2012poincare,rosales2018review}. 

\begin{figure*}
\centering\includegraphics[width=\linewidth]{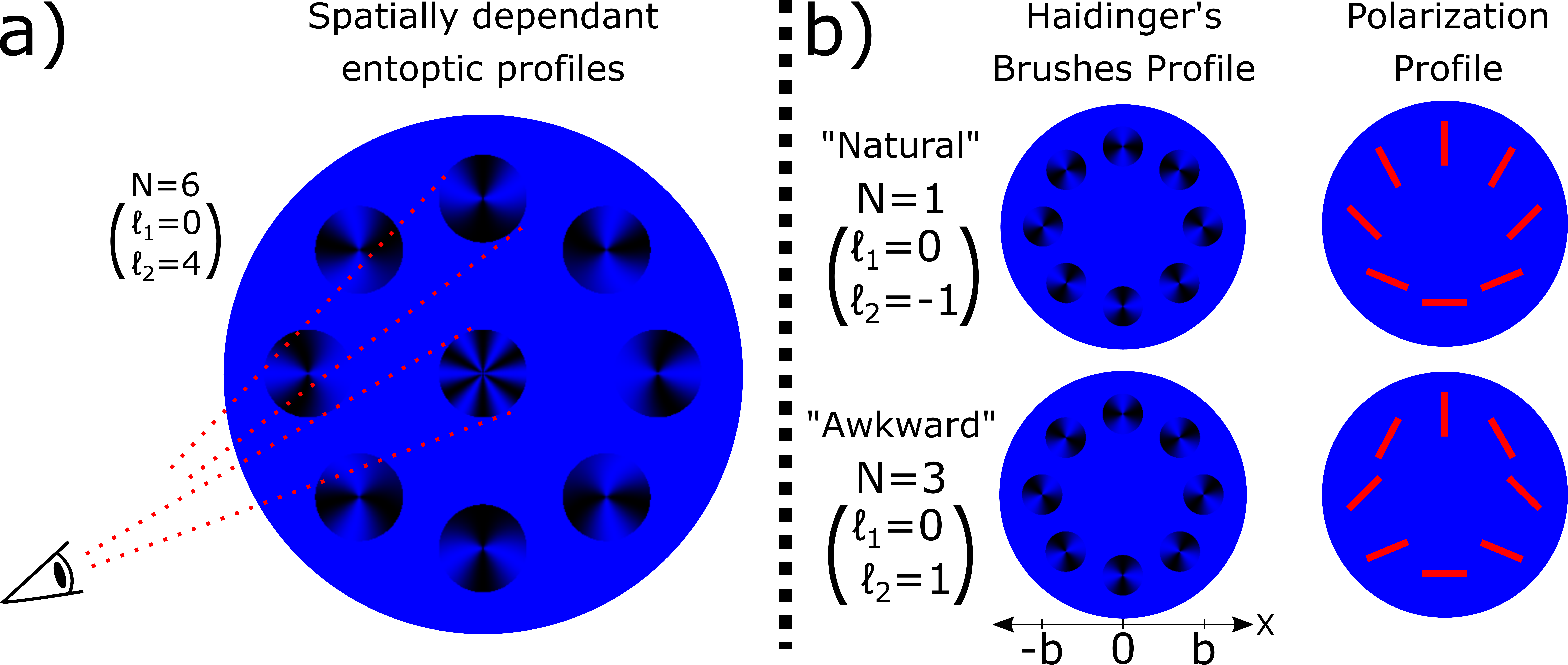}
\caption{a) Pictorial representation of the spatially dependant entoptic profiles that a human observer would perceive when directly viewing different regions of a spin-orbit beam described by Eq.~\ref{Eqn:PsiInGeneral} where $\ell_1=0$ and $\ell_2=4$.  
If the observer looks at the polarization gradient in the vicinity of the center  they will perceive $|(\ell_1-\ell_2)-2|=6$ azimuthal fringes. If the observer looks far away from the center they will observe specifically oriented Haidinger's brushes as the polarization is roughly uniform in the field of vision. b)  In this study the strong polarization gradients in the central region of the beam were reduced by preparing spin-orbit states with radially dependent terms, as described by Eq.~\ref{Eqn:psipm}. The beam state was not varied with time, and the participants were asked to move their gaze around the perimeter of the beam while observing and discriminating the elicited Haidinger's brush profile, labelled either the ``awkward'' or ``natural'' profile depending on whether the entoptic brush appeared to rotate along or against the oberver's eye movement. In order to minimize visual adaptation to optical polarization the suggested speed for the self-generated circular eye motion was $1$ Hz. 
}
 \label{fig:fig1}
\end{figure*}

Due to their unique propagation properties spin-orbit beams have found numerous applications in high-resolution imaging, communication protocols, and optical metrology~\cite{rubinsztein2016roadmap, milione20154, marrucci2011spin}. Ref.~\cite{sarenac2020direct} extended these applications to visual science which sees a growing interest in integrating human detectors with recent technological advances ~\cite{tinsley2016direct,loulakis2017quantum,sim2012measurement,dodel2017proposal,margaritakis2020spatially}. It was shown that humans are able to perceive and discriminate spin-orbit states through entoptic images that arise from the interaction between the 2D polarization topologies of these beams and the radially symmetric dichroic elements in the macula of the human eye~\cite{sarenac2020direct}. As shown in Fig.~\ref{fig:fig1}a, when looking in the vicinity of the center of a spin-orbit beam composed of a superposition of right and left circular polarization coupled to two different OAM values ($\ell_1$ and $\ell_2$), the observer may perceive an entoptic profile composed of $N=|(\ell_1-\ell_2)-2|$ azimuthal fringes. Ref.~\cite{sarenac2020direct} employed refractive elements to generate OAM=7 and couple it to different polarization states to induce $N=5$ and $N=9$ azimuthal fringes. Here we consider spin-orbit states prepared through devices that manipulate geometric phases. This enables a robust study of discriminating between beams with different OAM values.

The detection of optical polarization by humans is enabled by a series of radially symmetric dichroic elements centred on the foveola in the human eye~\cite{haidinger1844ueber,misson2015human,misson2017spectral,misson2019computational,horvath2004polarized}. An observer directly viewing polarized light would perceive a bowtie-like shape, known as ``Haidinger's Brush'', in the central point of their visual field. These entoptic profiles can be observed, on average, in light with more than approximately $56\%$ polarization~\cite{temple2015perceiving}, with peak clarity occurring for blue light of approximately $460$ nm wavelength~\cite{bone1980role}. However, observation of Haidinger's Brush in everyday life is hindered by visual adaptation which causes the entoptic images to disappear after a few seconds. Experiments show that a rotating polarization source at approximately $1$ Hz allows Haidinger's Brush to be observed continuously with clarity~\cite{coren1971use}.

Typical studies with human detectors of optical polarization require the observer to keep their gaze on a fixed point while the beam is time modulated to overcome visual adaptation. Correspondingly, in Ref.~\cite{sarenac2020direct} an induced phase shift caused the entoptic profile to rotate. In the presented study we consider an alternative method of observation utilizing stationary beams. Rather than fixing their gaze at the center of the beam, the participants completed a circular eye movement and determined which state they were presented with based on the rotational symmetry of the Haidinger brushes (see Fig.~\ref{fig:fig1}b).

The transverse wavefunction of a spin-orbit state travelling along the z-direction can be written as:

\begin{align}
	\ket{\Psi}=\frac{1}{\sqrt{2}}\left[C_1(\ell_1,r,z)e^{  i\ell_1\phi}\ket{R}
                +C_2(\ell_2,r,z)e^{  i\ell_2\phi}\ket{L}\right],
	\label{Eqn:PsiInGeneral}
\end{align}

\noindent where we have used the bra-ket notation for convenience, $\ket{L}=\begin{pmatrix}0 \\ 1 \\ \end{pmatrix}$ and $\ket{R}=\begin{pmatrix}1 \\ 0 \\ \end{pmatrix}$ denote the left and right circular polarization, and $(r,\phi)$ are the cylindrical coordinates. The radial terms $C_1(\ell_1,r,z)$ and $C_2(\ell_2,r,z)$ depend on the preparation method~\cite{LesAllen1992,berry2004optical}.

In our setup the spin-orbit states were prepared via Lattice of Optical Vortices (LOV) prism pairs which induce highly uniform phase gradients and minimize distortions in the beam's intensity profile~\cite{sarenac2018generation,schwarz2020talbot,sarenac2018methods}. Furthermore, they introduce a radial dependence which can be used to prepare a cue that helps participants learn the proper eye movement. In the study we specifically prepare and differentiate between the following two states:

\begin{figure*}
\centering\includegraphics[width=\linewidth]{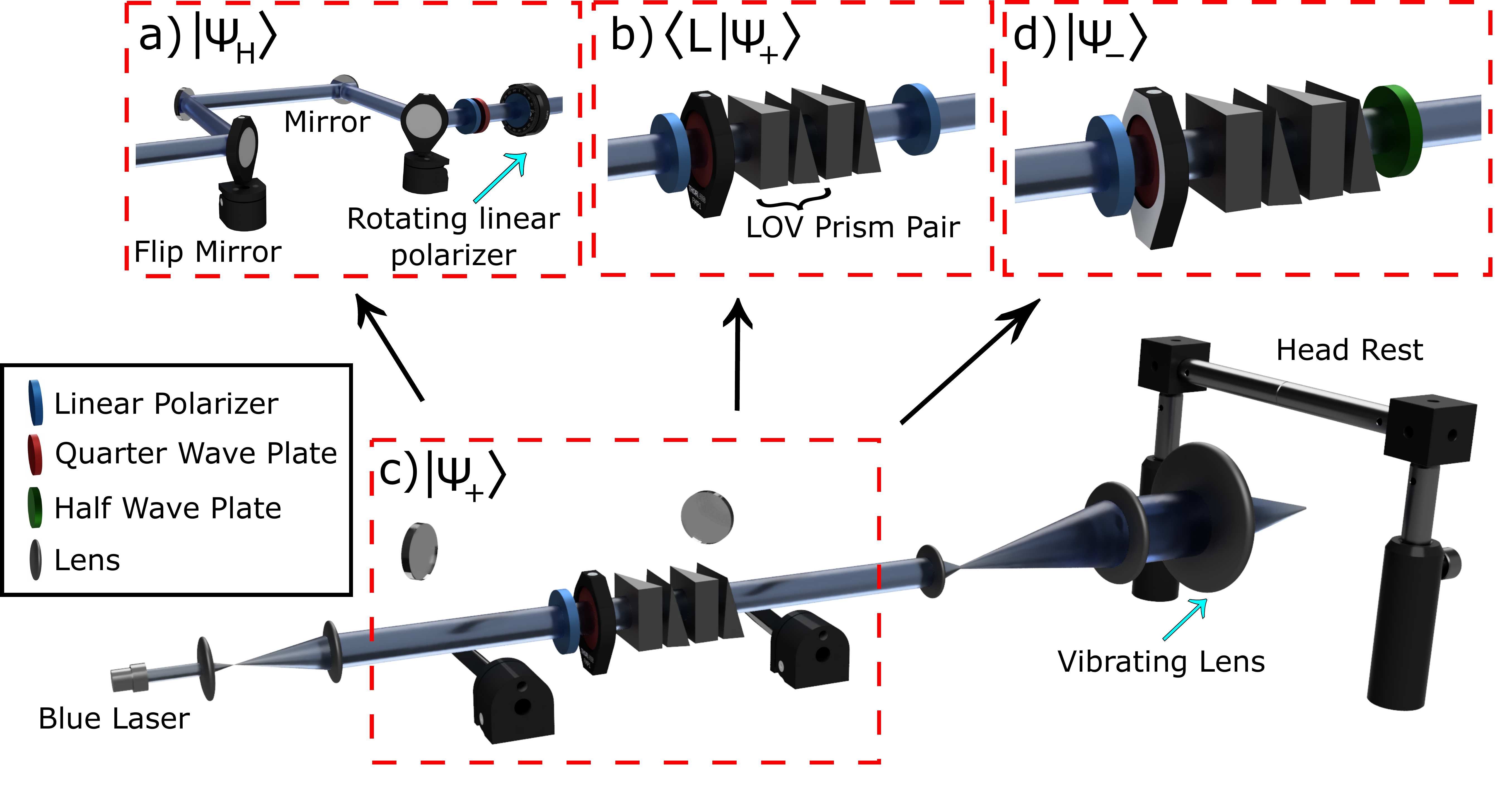}
\caption{Schematic of the experimental setup with four different configurations for directing structured light onto the retina of the observer. a) The first configuration prepares linearly polarized light with an orientation that rotates with a speed of approximately $1$ Hz. This setup was used in the pre-study to determine if the participants were able to perceive Haidinger's brushes when gazing at a fixed point. b) The second configuration prepares light with an intensity profile of $|\braket{L|\Psi_+}|^2$, where $\ket{\Psi_+}$ is given by Eq.~\ref{Eqn:psipm}. The doughnut shaped intensity profile defined by the $\sin^2(\pi r/4b)$ term allows for the identification of the circle of radius $b$ (see Fig.~\ref{fig:fig1}b) to be traced out during the eye movement task. The step was used to familiarize the participant with the size and location of the circle. The last two configurations, c) and d), prepare the two states of Eq.~\ref{Eqn:psipm}.
}
 \label{fig:setup}
\end{figure*}

\begin{align}
	\ket{\Psi_\pm}\approx\cos\left(\frac{\pi r}{4b}\right) \ket{R}
               \pm i\sin\left(\frac{\pi r}{4b}\right)e^{\pm i\phi}\ket{L},
	\label{Eqn:psipm}
\end{align}
\noindent where $b$ is the distance at which there is an equal amount of the two circular polarization states. The helical phase denoting the OAM is induced via the Pancharatnam-Berry phases. It follows from Eq.~\ref{Eqn:psipm} that the left circular polarization state of $\ket{\Psi_+}$ carries an OAM of +$1$ and that of $\ket{\Psi_-}$ carries an OAM of $-1$. Polarization directions, and the orientations of the corresponding Haidinger brushes, are shown in  Fig.~\ref{fig:fig1}b for several regions centered on $r=b$. Note that the polarization topologies of $\ket{\Psi_+}$ and $\ket{\Psi_-}$ depict the ``star'' and ``lemon'' Poincar\'e beams~\cite{zhan2009cylindrical,cardano2012polarization,galvez2012poincare,rosales2018review}.

The OAM dynamics were negligible and the beams possessed highly uniform intensity profiles in our setup. As a result, a speckle pattern arose which greatly hindered the observation of entoptic images. Therefore, we employed a vibrating lens, which is a common method of removing speckle patterns~\cite{Furukawa2008Speckle}. 

In typical operation two sets of LOV prism pairs take as input a circularly polarized state and output a lattice of spin-orbit states. The parameters in the experiment were such that one unit cell covered the entire beam, and the state in each unit cell of the lattice depended on the polarization of the input state as follows:

\begin{align}
	\ket{R}&\rightarrow \cos\left(\frac{\pi r}{4b}\right) \ket{R}
                +i\sin\left(\frac{\pi r}{4b}\right)e^{ i\phi}\ket{L}\\
	\ket{L}&\rightarrow \cos\left(\frac{\pi r}{4b}\right) \ket{L}
                +i\sin\left(\frac{\pi r}{4b}\right)e^{- i\phi}\ket{R}
	\label{Eqn:LOVmapping}
\end{align}

\noindent It follows that $\ket{\Psi_+}$ may be obtained by passing $\ket{R}$ through two sets of LOV prism pairs, while to obtain $\ket{\Psi_-}$ we start with an input of $\ket{L}$ and add a half wave plate at the end.

Our setup allowed for the robust and quick access to four different configurations, as shown in Fig.~\ref{fig:setup}. For a detailed description of the setup see Appendix A. The first configuration, Fig.~\ref{fig:setup}a, prepares a uniform intensity beam with a varying linear polarization direction:

\begin{align}
	\ket{\Psi_\text{H}}\approx\ket{R}
                +e^{i\theta (t)}\ket{L},
	\label{Eqn:psi}
\end{align}

\noindent where $\theta (t)$ was set by the rotation stage onto which a polarizer was mounted. This configuration was used in the study for pre-screening of the participants. The second configuration prepares the $\ket{\Psi_+}$ state (described by Eq.~\ref{Eqn:psipm}) filtered on $\ket{L}$. The resulting intensity possesses a doughnut profile that is described by $\sin^2\left(\pi r/4b\right)$. The outer boundary of the central dark region provides an aligning aid for the participant as it outlines the circle that should be traced by the participant's eye movements when viewing the spin-orbit states. The third and fourth configurations are the $\ket{\Psi_+}$ and $\ket{\Psi_-}$ spin-orbit states as described in Eq.~\ref{Eqn:psipm}. 

\begin{figure}
\centering\includegraphics[width=\linewidth]{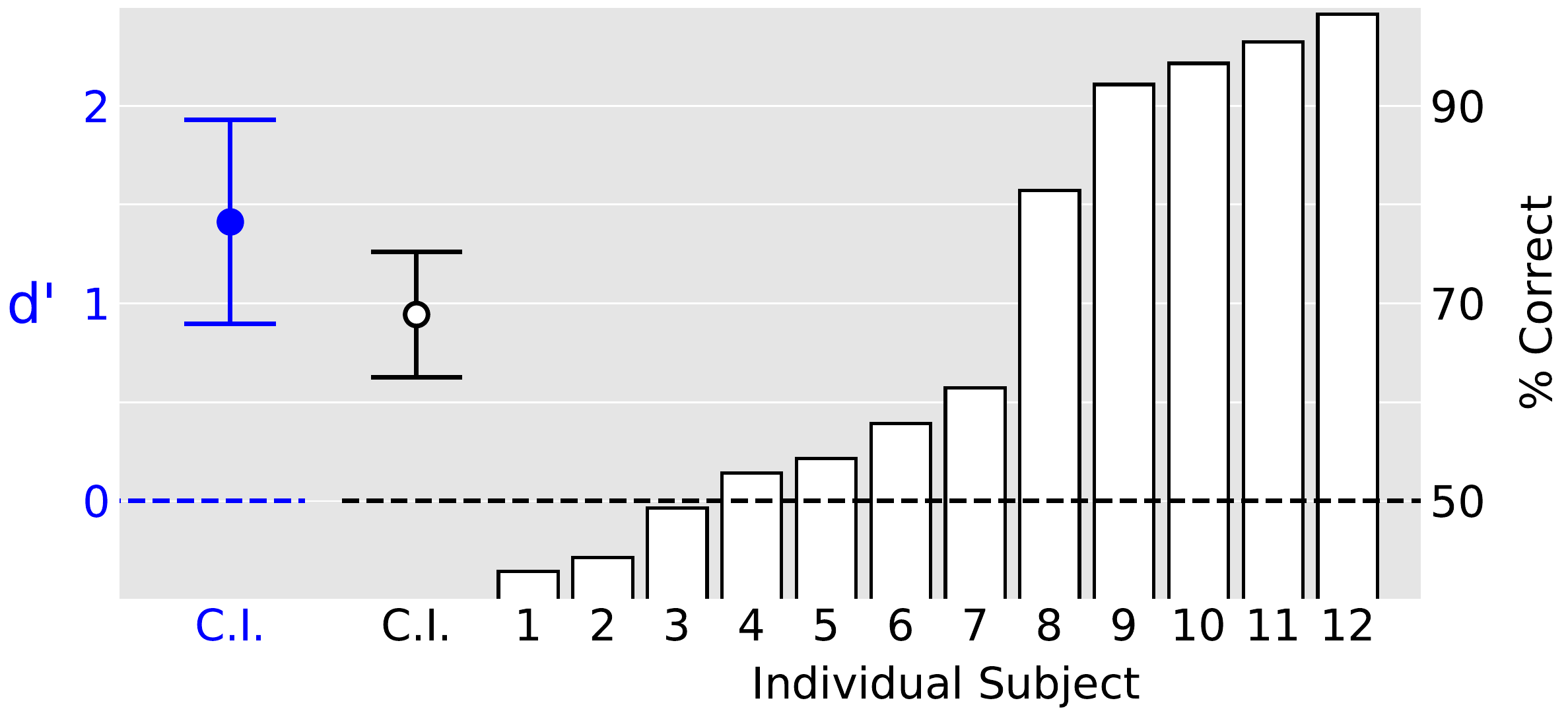}
\caption{Sensitivity and accuracy for the discrimination task. Each participant performed 140 trials over four daily sessions. The dashed lines indicate chance performance. The open bars illustrate individual rank-ordered performance, and the circular symbols illustrate group mean sensitivity (blue: left ordinate) and accuracy (black: right ordinate). The error bars are 95 \% confidence intervals. Generally, participants either performed very well or at chance levels.}
 \label{fig:results}
\end{figure}

After an initial screening and familiarization session, participants performed the discrimination task over four sessions, each containing thirty-five trials. There was no statistical difference between the results across days, and therefore the data from all sessions were collapsed for the main analysis. Sensitivity ($d^\prime$) and response bias ($c$) were calculated for each individual participant~\cite{wickens2002elementary}. Accuracy (\% correct), sensitivity, and response bias were analyzed using two-tailed one-sample \textit{t}-tests. Participants performed significantly better than chance: Accuracy: 69\% SD = 22, \textit{t}(11) = 3.0, \textit{p} = 0.013; Sensitivity $d^\prime$: 1.4, SD = 1.8, \textit{t}(11) = 2.7, \textit{p} = 0.019. No significant response bias was observed, $c$ = 0.028, SD = .17, \textit{t}(11) = 0.6, \textit{p} = 0.6. Fig.~\ref{fig:results} illustrates group and single-subject data.

Inspection of Fig.~\ref{fig:results} reveals a striking bimodal distribution of task performance with subjects 1 to 7 performing at chance levels and subjects 8 to 12 performing at ceiling. One explanation for this pattern of results is that seven participants were unable to perform the relatively complex psychophysical task that required the combination of self-generated eye movements with judgements of Haidinger’s brush rotation direction. However, it is also possible that the bimodal distribution reflects individual differences in macular pigment structure. Initial support for this intriguing possibility can be found in Ref.~\cite{sarenac2020direct} where the same bimodal task performance distribution was observed for a much easier psychophysical task that employed a rotating beam and did not require self-generated eye movements. Fig.~\ref{fig:ComparisonResults} in the appendix provides a visual comparison of both datasets for the six participants that completed both studies and confirms the robust nature of the bimodal performance distribution. Our future studies will aim to gain further insights into the functional properties of individual retinal pigment variations by examining the relationship between psychophysical performance and retinal images.

While the task employed in Ref.~\cite{sarenac2020direct} may provide high resolution for capturing the full spectrum of spin-coupled OAM sensitivity, the current task appears exceedingly effective at identifying high and low spin-coupled OAM discrimination performers, such that anybody unable to achieve ceiling performance will invariably perform near chance. Future work should apply these structured light discrimination tasks to measure properties of the macular pigment, allowing meaningful applications for diseases affecting the macular pigment such as macular degeneration~\cite{forster1954clinical,naylor1955measurement,muller2016perception}.

\section{Acknowledgements}

This work was supported by the Canadian Excellence Research Chairs (CERC) program, the Natural Sciences and Engineering Research Council of Canada (NSERC) grants [RGPIN$-2018-04989$], [RPIN$-05394$], [RGPAS$-477166$], the Collaborative Research and Training Experience (CREATE) program, the Government of Canada’s New Frontiers in Research Fund (NFRF) [NFRFE$-2019-00446$], the Velux Stiftung Foundation [Grant 1188], and the Canada  First  Research  Excellence  Fund  (CFREF).

\bibliography{OAM}

\clearpage

\section{APPENDIX}
\subsection{Setup and Stimuli}
The setup used a laser of wavelength 455 nm and the beam was attenuated to $\textless1$ $\mu$W/mm$^2$ at the location of the observer in order to conform to the guidelines for laser exposure time outlined by the International Commission on Non-Ionizing Radiation ~\cite{international2000revision}. The beam was passed through a single mode fiber followed by two lenses which expanded the beam to a diameter of approximately $2$  cm. 

There are four different configurations in the setup. The common parts that are present in each configuration are the mentioned fiber, attenuators, and lenses at the first stage of the setup, and then also the parts in the last stage of the setup that include the lenses used to expand the beam to a diameter of approximately $2$ inches and the user lens  ($f=150$ mm) that directed the beam onto the retina of the observer. To remove the speckle pattern the user lens was vibrated via push-pull motor that operated at 60 Hz with a stroke of 5 mm along the direction perpendicular to the beam propagation axis. Finally, a headrest, that included a chin rest with a variable height and a forehead rest bar, was placed at the end of the setup. The location of the headrest was optimized for each participant. The participants covered their non-viewing eye with an eye patch. 

The first configuration of the setup (see Fig.~\ref{fig:setup}a) uses mirrors to redirect the beam around the main optical components. A linear polarizer, quarter wave plate, and a rotating polarizer are then used to generate light whose polarization direction rotates in time.

The next three configurations use two sets of LOV prisms to generate the desired spin-orbit beams. The LOV prism pairs were circular quartz wedges with a wedge angle of $2\degree$, a diameter of $2.54$ cm, and for one wedge the optical axis was aligned with wedge angle while for the other wedge the optical axis was aligned $45\degree$ to wedge angle. Their characteristics are described in detail in Ref.~\cite{sarenac2018generation} and Ref.~\cite{schwarz2020talbot}.

The participants were tasked with discriminating between $\ket{\Psi_+}$ and $\ket{\Psi_-}$ based on the rotational symmetry of the Haidinger's brushes along the circle of radius $b$. Because the direction of rotation of the Haidinger's brushes of $\ket{\Psi_+}$ ($\ket{\Psi_-}$) is the same (opposite) relative to the direction of the eye movement, this trial type was given the label ``natural'' (``awkward''). 

\subsection{Participants}
Experimental participants were recruited from the Institute for Quantum Computing and the School of Optometry and Vision Science at the University of Waterloo. The complete study involved five experimental sessions. All research procedures received approval from the University of Waterloo Office of Research Ethics and all participants were treated in accordance with the Declaration of Helsinki.

A total of 17 participants, two of which are authors on the current study, were recruited. Participation was entirely voluntary and all participants, aside from the authors, received \$15 CAD per session in appreciation for their time. Out of the 17 participants, 5 participants were unable to pass the screening on the second day and their participation was terminated. Of the 5 who did not complete the study, 4 were unable to see Haidinger's brushes with the setup in the first configuration, and 1 was unable to perform the physical task without head movement. Therefore, a total of 12 participants completed the experiment.

\subsection{Psychophysical Procedure}
Participants were trained and tested on a psychophysical discrimination task over five experimental sessions. During Session 1, participants were introduced to the concepts relating to the human perception of polarization, and the task was explained. The participants viewed animated diagrams of natural and awkward trials. The participants were then presented with the first configuration of the setup (see Fig.~\ref{fig:setup}a). The polarization of the light rotated either clockwise or counterclockwise, and participants were asked to fixate at the center of the beam and indicate the direction of rotation of the Haidinger's brush. Finally, the second and third configurations of the setup (see Fig.~\ref{fig:setup}b$\&$c) were presented to the participant, and they were trained to execute the correct eye movement - eye movements along the indicated circle within the beam while keeping the head still - while observing the spin-orbit beam.

During Session 2, participants completed an assessment before beginning the main task. They were again exposed to the the first configuration of the setup (see Fig.~\ref{fig:setup}a) and performed the orientation discrimination task. The second configuration of the setup (see Fig.~\ref{fig:setup}b)  was then presented and the ability to perform the correct motor action was assessed. Participants were excluded from further testing if they could not identify the direction of the Haidinger's brush with a $\textgreater70\%$ 
probability or if they produced visually noticeable head movement while viewing the beam from the second  configuration and attempting the self directed eye motion. Participants who successfully completed the assessment were then able to perform the main task.

Participants performed the main psychophysical task during Sessions 2, 3, 4, and 5, following an identical procedure. During testing, all participants observed the beam with their preferred eye and the other eye was occluded. Each session was composed of five blocks. At the start of each block, participants observed two alternating presentations of "awkward" and "natural" stimuli, and participants observed the stimuli by freely generating the correct eye movement for 15 seconds. After this free-viewing period, the actual discrimination task began. Seven trials were presented per block, each trial presenting either an "awkward" or a "natural" stimulus for 45 seconds, and the participant verbally indicated the perceived trial type. Real-time corrective feedback was given. Each block contained 7 trials, and each participant completed 5 total blocks per session. In total, 35 trials were completed per session and each participant completed a total of 140 trials across 4 testing sessions. Participants were given a 1 minute break in between blocks.

\subsection{Comparison to previous study}

The following figure provides a visual comparison of both datasets for the common participants of the presented study and that of Ref.~\cite{sarenac2020direct}.

\begin{figure}
\centering\includegraphics[width=\linewidth]{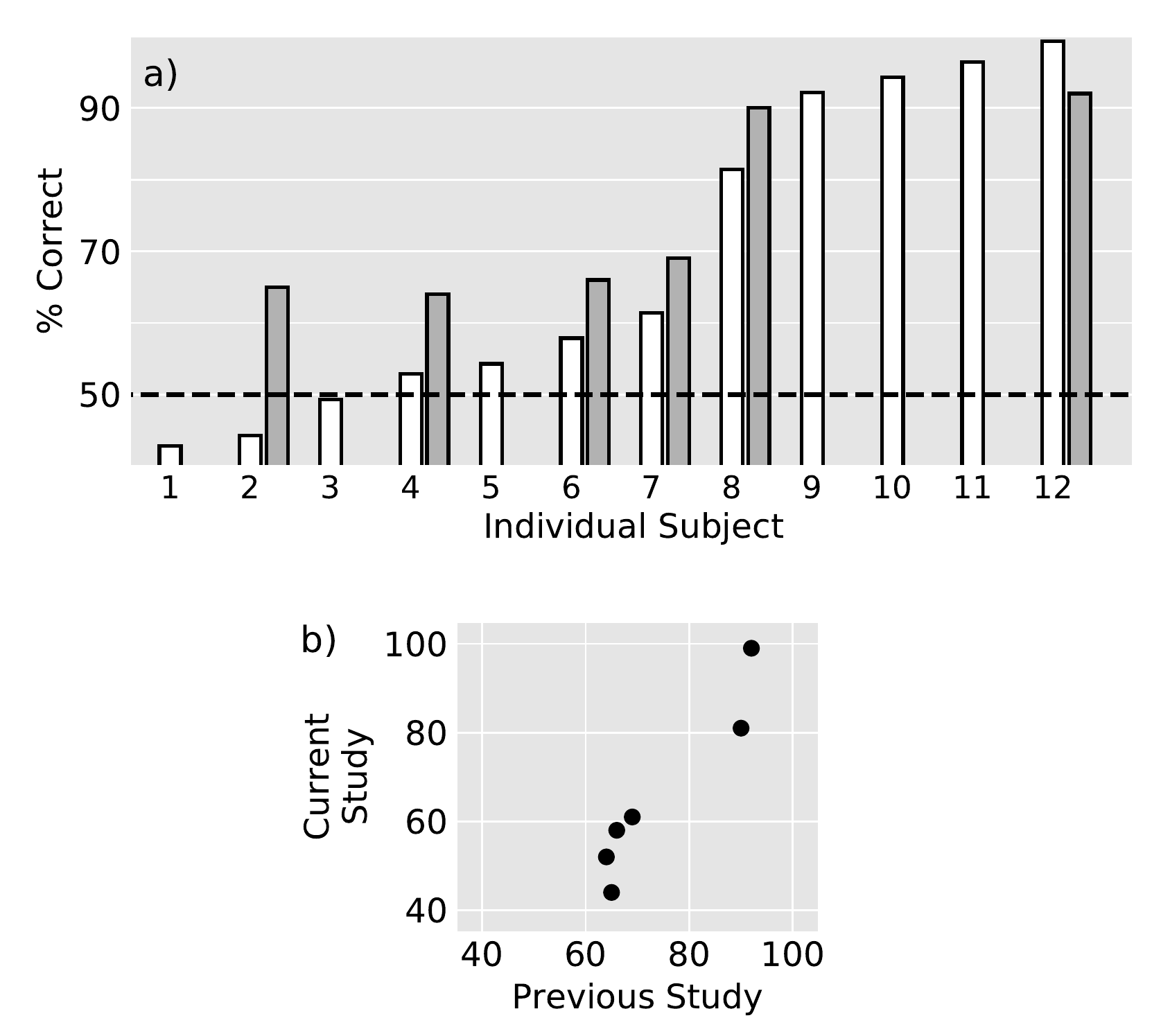}
\caption{a) Results of the current experiment (white bars) plotted alongside results of Ref.~\cite{sarenac2020direct} (gray bars) for overlapping participants. b) Single-subject results (y-axis: current study) plotted against  the same subject's performance in the previous study Ref.~\cite{sarenac2020direct} (x-axis).}

\label{fig:ComparisonResults}
\end{figure}

\end{document}